\documentclass[10pt, conference, letterpaper]{IEEEtran}
\IEEEoverridecommandlockouts
\usepackage{cite}
\usepackage{amsmath,amssymb,amsfonts, amsthm}
\usepackage{algorithmic}
\usepackage{graphicx}
\usepackage{float}
\usepackage{subfig}
\usepackage{textcomp}
\usepackage{xcolor}
\usepackage{multirow}
\usepackage{placeins}
\usepackage{hhline}
\usepackage{tabularx}
\usepackage{enumitem}
\usepackage{setspace}
\usepackage{booktabs}
\usepackage{adjustbox}
\usepackage{lipsum}
\usepackage{mathtools}
\usepackage{cuted}
\usepackage[normalem]{ulem}
\usepackage[ruled,vlined]{algorithm2e}
\include{pythonlisting}
\newcommand*\diff{\mathop{}\!\mathrm{d}}

\SetKwInput{KwInput}{Input}                
\SetKwInput{KwOutput}{Output}              

\DeclareMathOperator*{\reply}{reply}
\DeclareMathOperator*{\main}{main}
\DeclareMathOperator*{\MAE}{MAE}

\makeatletter
\def\endthebibliography{%
	\def\@noitemerr{\@latex@warning{Empty `thebibliography' environment}}%
	\endlist
}
\makeatother

\def\BibTeX{{\rm B\kern-.05em{\sc i\kern-.025em b}\kern-.08em
		T\kern-.1667em\lower.7ex\hbox{E}\kern-.125emX}}
\begin{document}
	\newcolumntype{C}[1]{>{\centering\arraybackslash}p{#1}}
	\title{NesTPP: Modeling Thread Dynamics in Online Discussion Forums\\}
	
	\author{\IEEEauthorblockN{
			Chen Ling\IEEEauthorrefmark{1},
			Guangmo Tong\IEEEauthorrefmark{1}, and
			Mozi Chen\IEEEauthorrefmark{2}}\\
		\IEEEauthorblockA{\IEEEauthorrefmark{1}Department of Computer and Information Sciences, University of Delaware, USA\\}
		\IEEEauthorblockA{\IEEEauthorrefmark{2}Department of Computer Science, Wuhan University of Technology, China} 
		\IEEEauthorblockA{\{lingchen, amotong\}@udel.edu, chenmz@whut.edu.cn}
	}
	
	\maketitle
	
	\begin{abstract}
		Online discussion forum creates an asynchronous conversation environment for online users to exchange ideas and share opinions through a unique thread-reply communication mode. Accurately modeling information dynamics under such a mode is important, as it provides a means of mining latent spread patterns and understanding user behaviors. In this paper, we design a novel temporal point process model to characterize information cascades in online discussion forums. The proposed model views the entire event space as a nested structure composed of main thread streams and their linked reply streams, and it explicitly models the correlations between these two types of streams through their intensity functions. Leveraging the Reddit data, we examine the performance of the designed model in different applications and compare it with other popular methods. The experimental results have shown that our model can produce competitive results, and it outperforms state-of-the-art methods in most cases.
	\end{abstract}
	
	\begin{IEEEkeywords}
		temporal point process, online discussion forum, information dynamics
	\end{IEEEkeywords}
	
	\section{Introduction}
	
	
	The ways of human interaction have long been a subject of interest. Recently, the rise of online social networks provides a rich substrate for large-scale online user interaction for sharing and exchanging information. Understanding the latent mechanism of such information dynamics has been an active topic, because it provides us with new insights into the needs, opinions, and experiences of individuals \cite{smedley2018practical}. Through modeling such information diffusion dynamics, the existing works have greatly promoted the development of online social networks by various applications, involving content recommendation (\hspace{1sp}\cite{asur2010predicting, lan2018personalized}): a high-quality
    post can be recommended by certain standards, and anomaly information and misinformation detection (\hspace{1sp}\cite{wu2016mining, rizoiu2017expecting, liu2019latent}).
	

    As a subclass of online social networks, the online discussion forum (ODF) has become a significant platform of knowledge sharing for the past decade. In the area of online education, MOOC (Massive Online Open Course) is an exemplary ODF that enables students in online courses to learn socially as a supplement to their studying of the course content individually. According to Alexa.com\footnote{https://www.alexa.com/topsites} by Amazon,  Reddit is now the second-largest social networking site in terms of the number of daily visitors and previews. ODFs are distinct from other platforms in primarily two aspects. First, unlike Facebook or Twitter, ODFs are not built through user relationships (e.g., friendship and follower-followee relationship), and their contents are publicly available to all visitors. Online users have the flexibility to reflect on their thoughts and read the responses of any public posts. Second, ODFs allow high-volume asynchronous interactions through the thread-reply mode in which participants can create threads (e.g., posts and questions) and react to the threads by making replies (e.g., comments, votes, and answers)\cite{im2006online}. Universally, the ODF promotes users to popularize well-regarded threads to the front page via replies, and such a process of information exchange exhibits a unique nested thread-reply structure in which these two types of streams are correlated with each other. Given these unique characteristics and successful examples, we in this paper study the problem of modeling thread dynamics in ODFs.

	\textbf{Challenges.} The primary problem in modeling information dynamics is to predict the arrival time of the subsequent events (\hspace{1sp}\cite{asur2010predicting, zhao2015seismic, kobayashi2016tideh, mei2017neural, chen2018marked}). There are two main challenges in studying the thread dynamics in ODFs. First, we often have very limited historical temporal information (e.g., arrival time or inter-arrival time) that is not sufficient for learning an expressive model, which is exacerbated by the high uncertainty of the eventual scope of an information cascade \cite{cheng2014can}. The common solution is to incorporate exogenous features (e.g., context, user profile, and other network information) to support the prediction. However, discriminative models with extra features cannot make adaptive predictions as the features of future events are not known to us. In the scenario of ODFs, if we would use context features, it is not possible to predict the arrival time of future threads adaptively unless we could also predict the future contents. On the other hand, training a generative model that simultaneously predicts all the features can be overambitious since we, in fact, care only the temporal patterns but not everything happening in the future. Secondly, despite the extensive research carried out on analyzing the unique discussion structure of the ODF (\hspace{1sp}\cite{thomas2002learning, aumayr2011reconstruction, medvedev2018modelling}), few studies have considered if there exists an impact of the natural thread-reply structure to the information exchange pattern in the ODF. Moreover, can we utilize the unique nested structure of ODFs in predicting its future thread dynamics? In this work, we aim to propose an effective framework to resolve such difficulties.

	\textbf{Our Work.} In order to model thread dynamics in ODFs, we design the \underline{Nes}ted \underline{T}emporal \underline{P}oint \underline{P}rocess (NesTPP) model that leverages the unique nested thread-reply structure without using excessive features. The major contributions are summarized as follows. 

	\begin{itemize}
		\item We build the NesTPP framework for modeling temporal dynamics in the ODF by taking the main thread stream and associated reply stream as combined self-exciting temporal point processes with carefully designed intensity functions. 
		
		\item We derive the closed-form likelihood function of NesTPP, and an MLE-based algorithm, as well as a quasi-Newton optimization method, are utilized to estimate the parameters in the combined self-exciting point process. An adaptive sampling algorithm is also proposed to simultaneously generate the future main threads and replies without assuming a fixed dimension, making NesTPP applicable to a few applications. 
		
		\item We show that NesTPP is practically effective through the experiments on real-world social media data. According to the results, the NesTPP model produces a clear improvement over the existing state-of-the-art methods in several prediction tasks. In addition, we demonstrate that NesTPP can be used as a tool for visualizing the correlation between event streams.
	\end{itemize}

	\textbf{Road-map.} Sec. \ref{sec: related_work} surveys the related works. In Sec. \ref{sec: model}, we describe the NesTPP model and its properties. The following Sec. \ref{sec: experiment} presents the experiment results. In the end, Sec. \ref{sec: conclusion} concludes the paper and discusses potential future works.
	
	\section{Related Work} \label{sec: related_work}
	\textbf{Temporal Point Process (TPP).} Temporal point processes leverage the fine-grained time information and related features from the event stream to understand the diffusion mechanism in social networks, which have been viewed as a natural choice of modeling social interactions for many years. Among the variants, Hawkes process \cite{hawkes1971point}, as a subset of non-homogeneous Poisson process, has been broadly employed in modeling more intricate information propagation process. Zhao \textit{et al.} \cite{zhao2015seismic} implemented the SEISMIC model based on the self-exciting point process to predict the final volume of the retweet cascade. Similarly, Kobayashi \cite{kobayashi2016tideh} and Chen \cite{chen2018marked} utilized additional mark information\footnote{Features other than event arrival time, such as the number of followers of a user, the periodic factor, and the social popularity of a discussion event.} to more accurately model and predict the retweeting dynamics in Twitter. There also exists literature \cite{iwata2013discovering, linderman2014discovering, blundell2012modelling} studying the latent relationship between different components, involving users, social activities, and network structures, by the TPP model. NesTPP differs from the existing ones in that we regard both thread stream events and reply stream events as the separate but correlated self-exciting point processes in the same event space. Furthermore, our model utilizes only internally-generated mark information, enabling stronger applicability to other thread-reply scenarios.  Additionally, each thread-reply cascade in ODFs can be regarded as an individual point process, which forms a special case of the multivariate point processes with dynamic dimensions. Unlike other methods (\hspace{1sp}\cite{cao2017deephawkes, mei2017neural}) that adopted only fixed-dimension multivariate point processes, our model can adaptively learn the latent influence among thread-reply cascades and simulate new cascade by creating new dimensions. 
	
	\textbf{Online Discussion Forum.} In the study of ODFs, a large volume of works (\hspace{1sp}\cite{yang2003effects, dawson2008study, yusof2009students, lan2018personalized}) have investigated the information exchange dynamics in MOOC platform (i.e., Massive Open Online Courses), a special form of an ODF for education only. Studying the information exchange dynamics in a more general ODF is needed. Recently, Gomez \textit{et al.} presented a generative model \cite{gomez2011modeling} to capture the temporal evolution of the observed thread-reply structure. In \cite{wang2012user}, the authors presented a dynamic model that predicts the growth trends and structural properties of the online conversation threads; however, their works have limitations in processing real-world data and predicting future cascade size because of the mean-field nature of the adopted L\'{e}vy process model. In addition, Ryosuke \cite{nishi2016reply} and Medvedev \cite{medvedev2018modelling} also proposed different models to predict the temporal dynamics of main threads as a discussion tree. More recently, \cite{liu2019latent} utilized the Hawkes process to target the user and event clustering in the ODF. Nonetheless, their works have treated all events in an ODF as a unified point process. In practice, the information diffusion pattern between the main thread stream and reply stream should be governed by different point processes since they may have different diffusion patterns. Moreover, existing works \cite{lan2018personalized,aumayr2011reconstruction, medvedev2018modelling} have used additional mark information in their models for various prediction tasks in ODFs. However, the application of future cascade prediction requires long-range adaptively prediction, which is not applicable to utilize external mark information to enhance the prediction accuracy. The NesTPP proposed in this paper focuses on utilizing only the natural nested structure and easy-to-retrieved mark information to capture more subtle shifts in the information cascade evolution.

	\textbf{Other Related Works.} Other than the point process based methods, there also exist a large volume of feature-based approaches (\hspace{1sp}\cite{bakshy2011everyone, cheng2014can, naveed2011bad, zaman2010predicting}) that study the information propagation pattern mostly in friendship-based social networks. These methods often need to extract extensive features and utilize learning algorithms (e.g., random forest, SVM, and neural networks) to capture the propagation pattern from data. Such approaches have limited applicability because feature engineering is often very expensive and time-consuming, and most of the works are classification-based models (e.g., predicting if an information cascade would go viral), which cannot be used to predict future cascade growth activity. As we stated in the Introduction, exogenous features are hard to retrieve in the task of adaptive prediction, and they cannot be applied directly in our scenario. Our work differs from them because the only feature utilized is the arrival time of each event; other mark information is also derived from the event arrival time to enhance the model's forecasting capability.

	\section{Nested Temporal Point Process: NesTPP}  \label{sec: model}
	In this section, we start by introducing the problem setting and the key concepts of the TPP. We then present the NesTPP model with the design of intensity functions, model complexity, sampling algorithm, and properties of our model.
	
	\subsection{Problem Setting} \label{subsec: setting}
	\begin{figure}[tbp]
		\centerline{\includegraphics[width=0.49\textwidth]{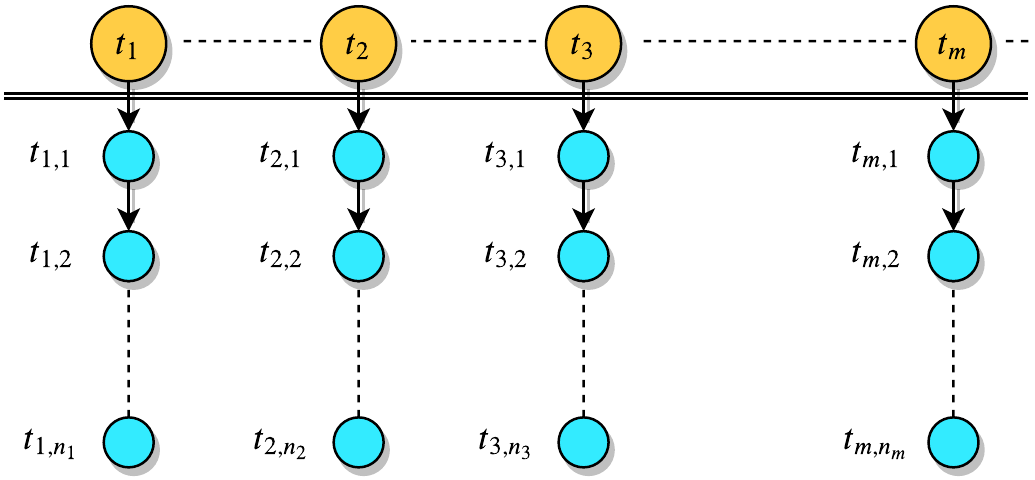}}
		\caption{The Nested Information Diffusion Structure in Forums}
		\label{fig:1}
	\end{figure}
	
	Our paper targets the online discussion forum operated in the \textit{thread-reply mode} composed of two types of event streams, \textit{thread stream} and \textit{reply stream}, where the threads  $\{t_i\}$ are created as new web content to which users can respond through replies $\{t_{i, j}\}$, as illustrated in Fig. \ref{fig:1}. Note that we use the arrival time of events to denote the events. In this paper, we aim to design a model that can understand the latent diffusion mechanism of the event streams under such a nested communication mode. Additionally, to better model the dynamics of information cascades in ODFs, most of the previous works utilized exogenous mark information to enhance the influence of past events in the point processes, including the number of up-votes, received awards, and the profile information of online users. However, additional mark information is usually hard to retrieve as it varies with time, and it thus cannot be directly employed to model the temporal change of the event cascade. 
	For better applicability, we utilize only the count of replies to the main thread at the current timestamp as the mark information. Table \ref{tab:my-table} summarizes the notations used in this work.
	
	\subsection{Temporal Point Process}
	In general, a TPP is defined as a random and finite series of events with associated marks governed by a probabilistic rule. Let a sequence of events $\{e_i\} = \{(t_i, p_i), i\in\mathbb{Z}^+\}$ represents a marked TPP, where $t_i \in \mathbb{R}^+$ is the occurrence time of each event and the $p_i$ denotes the corresponding event mark of $e_i$. We use $\mathcal{H}_t = \{t_i < t\}$ to denote the sequence of events arrival time until the current time $t$. The simplest class of TPP is the Poisson process in which events are arriving at an average rate of $\lambda \in \mathbb{R}^+$ per unit time, and the intensity of the Poisson process is defined as $\lambda$.
	
	A self-exciting point process (also known as the Hawkes Process) is essentially a non-homogeneous Poisson process where the event intensity $\lambda$ is determined by all previous events $\mathcal{H}_t$, and a conditional intensity function gives the expression of $\lambda$:
	\begin{equation}
	\label{eq: uni}
	\lambda (t\mid\mathcal{H}_t) = \mu_0(t) + \sum_{i:t>t_i}\psi(t-t_i, p_i),
	\end{equation} where $t \in \mathbb{R}^+$ represents the current timestamp, $ \mu_0(t) \in \mathbb{R}^+$ denotes the base intensity of each incoming event at time $t$, and $\psi(\cdot)$ is called the memory kernel. In specific, the memory kernel modulates the influence of a previous event at time $t_i$ to the later events. Typically, $\psi(\cdot)$ is taken to be monotonically decreasing so that more recent events have a higher influence on the current event intensity, compared to events having occurred further away in time \cite{rizoiu2017hawkes}. In this paper, we utilize the modified standard exponential decay memory kernel \cite{hawkes1971point} and power-law memory kernel \cite{ozaki1979maximum} to explicitly fit different event stream diffusion patterns.
	
	\begin{table}[t]
		\centering
		\begin{tabular}{c|l}
			\toprule
			Symbols          & \multicolumn{1}{c}{Meaning}                   \\ \midrule
			$t$ & The current timestamp                           \\
			$t_{i, j}$       & Arrival time of each event                         \\
			& $j=0$ denotes the arrival time of main thread \\
			$p_i$ & The number of replies of $i$-th main thread                           \\
			$\mathcal{H}_t$ & Historical events arrival time up to time $t$                           \\
			$\lambda(\cdot)$ & Intensity function                            \\
			$\psi(\cdot)$ & Memory kernel of NesTPP                            \\
			$\mu$ & Base intensity                            \\
			$\alpha$ & Triggering factor                            \\
			$\beta$ & Decaying rate of exponential decay function                           \\
			$\gamma$ & warping factor                            \\
			$c$ &  Regularization term                            \\
			$\eta$ & Decaying rate of power-law function                            \\
			$q(t)$           & Infectivity function of the linked main thread         \\ 
			$\delta$ & Decaying rate of infectivity function                             \\
			$n^*$ & Branching factor                             \\\bottomrule
		\end{tabular}
		\caption{Table of Notations}
		\label{tab:my-table}
	\end{table}
	
	%
	%

	\subsection{Nested Temporal Point Process: NesTPP} \label{sec:ntpp} 
\begin{figure}[tbp]
		\centerline{\includegraphics[width=0.48\textwidth]{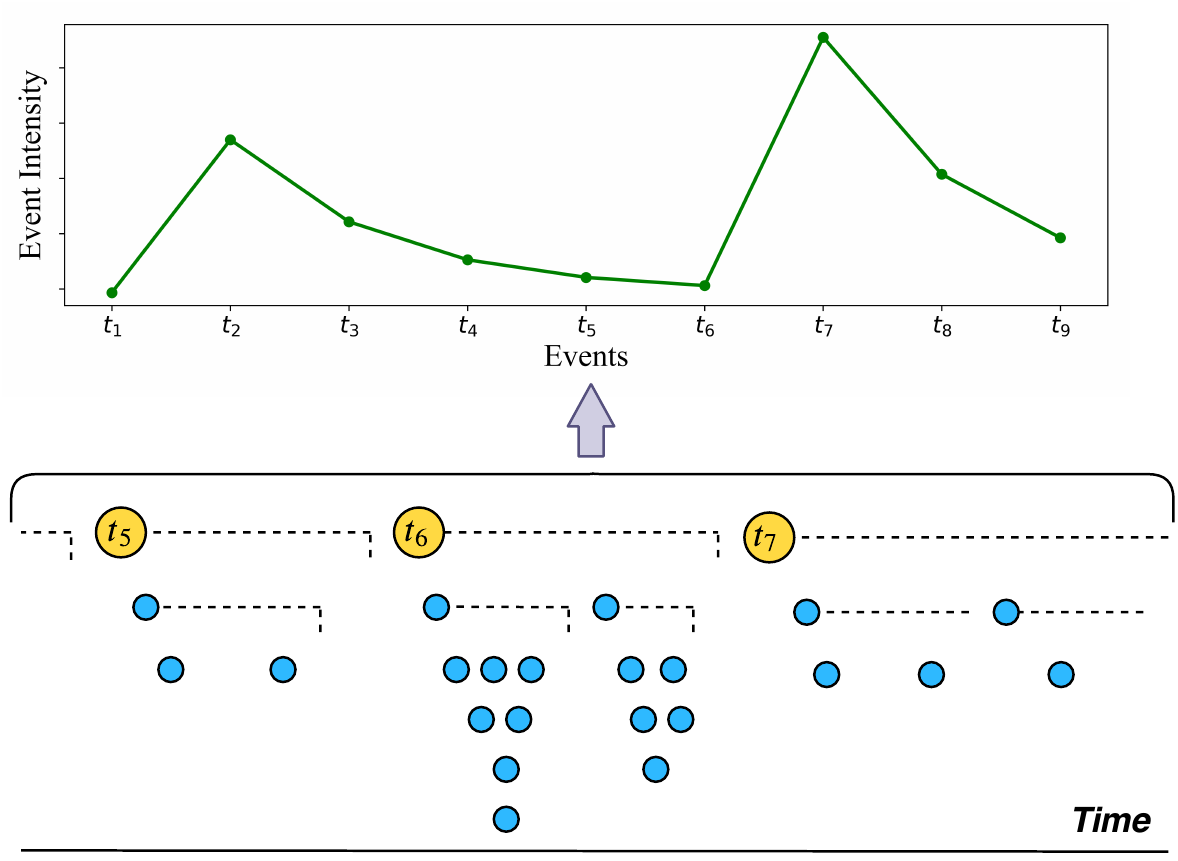}}
		\caption{The relation between main thread stream and reply stream simulated by NesTPP. The upper figure shows the evolution of event intensity, and the lower figure depicts a slice of the event pattern including both main threads and associated reply events. It is clear from the figure that thread $t_6$ draws tremendous replies and effectively stimulates the arrival of next thread $t_7$.}
		\label{fig:start}
		\vspace{-2mm}
	\end{figure}

	\textbf{Nested Structure.} The existing Hawkes-based point process models often treat the evolution of main threads and associated replies as a unified point process in one dimension. However, the diffusion patterns of different event streams are typically governed by different but correlated probabilistic rules. For example, the occurrence of a new main thread not only depends on the impact of all previous main threads but also from all the associated replies, and the generation of reply events also relies on the influence of the linked main threads. The latent relation between both event streams forms the nested structure of the information cascade. Fig. \ref{fig:start} illustratively presents the unique diffusion mechanism in an ODF. In order to comprehend such a mechanism, we design the NesTPP that treats both types of event streams as separate self-exciting point processes controlled by inter-dependent intensity functions. In what follows, we present detailed designs of our model.

	\textbf{Reply Stream.} The reply stream intensity function $\lambda_{\reply}^{i}(\cdot)$ derives from the intuition of the univariate Hawkes process defined in Eq. \ref{eq: uni} with an exponential decaying kernel. Recall that each main thread $t_i$ has $n_i$ reply events, and the arrival time of each linked reply is $t_{i, j}$. The intensity function of reply stream  $\lambda_{\reply}^{i}(\cdot)$ of $i$-th main thread is defined as: 
	\begin{equation}
	\label{eq: nest_reply}
	\lambda_{\reply}^{i}(t) = \mu_{\reply} + q(t) \cdot \sum^{n_i}_{j=1} \underbrace{\alpha e^{-\beta(t - t_{i, j})}}_{\psi_{\reply}(t)}.
	\end{equation} The term $\mu_{\reply} \in \mathbb{R}^+$ denotes the base intensity of the reply stream point process, and $\psi_{\reply}(t)$ represents the reply stream memory kernel. The positive constant $\alpha, \beta \in \mathbb{R}^+$ in the memory kernel have the following interpretations: each arrival of a reply in the system instantaneously increases the arrival intensity by $\alpha$, then the influence of this arrival decays at rate $\beta$ over time. Furthermore, $q(t) \in \mathbb{R}^+$ is the additional mark information which models the infectivity of the linked main thread, and it is defined as follows:
	\begin{equation*}
	\label{eq: infectiousness}
	q(t) = e^{-\delta (t - t_0)},
	\end{equation*} where $t - t_0$ denotes the duration of time from the current timestamp $t$ to the arrival time of the associated main thread $t_0$, and $\delta \in \mathbb{R}^+$ represents the decaying rate of the associated main thread infectivity to the reply stream. Intuitively, $q(t)$ scales down the estimated infectivity of the linked main thread over time, which accounts for the thread getting outdated based on the newsworthiness of a discussion topic.
	
	\textbf{Main Thread Stream.} Recalling the definition of the event intensity function in Eq. \ref{eq: uni}, the arrival rate of new threads depends on each previously occurred thread $t_i$ through the triggering kernel $\psi(t)$. In addition, the reply streams of each previously occurred thread also impact the arrival of new threads. We hereby construct the power-law based kernel function $\lambda_{\main}(t)$ as follows:
	\begin{equation}
	\label{eq: nest_main}
	\lambda_{\main}(t) = \mu_{\main} +  \sum_{i=1}^{m}\underbrace{\lambda_{\reply}^{i}(t) \cdot p_i^{\gamma}\cdot (t-t_i + c)^{-(\eta +1)}}_{\psi_{\main}(t)},
	\end{equation} where $\mu_{\main} \in \mathbb{R}^+$ is the base intensity of the main thread stream and $\psi_{\main}(t)$ is the memory kernel of main thread stream. $p_i^{\gamma} \in \mathbb{R}^+$ denotes the influence of the mark information from each individual main thread with $\gamma \in \mathbb{R}^+$ being a warping effect. Note that we assume that $p_i$ is the number of replies of the main thread up to the time $t_i$. Moreover, $(\eta +1), \eta \in \mathbb{R}^+$ is the power-law exponent representing the decaying rate of the influence of each previous event $t_i$, and parameter $c\in \mathbb{R}^+$ is a regularization term in order to keep the exponential term $\lambda_{\main}(\cdot)$ bounded when $t-t_i$ approaches $0$. Finally, the term $\lambda_{\reply}^{i}(t)$ denotes the influence factor carried by its associated reply streams, which has been defined in the Reply Stream part. In this power-law based intensity function, $\lambda_{\reply}^{i}(t)\cdot p_i^{\gamma} < \eta \cdot c^{\eta}$ denotes the overall magnitude of influence to the main thread stream. 
	
	In summary, the NesTPP models the cascade evolution by defining the TPP for each event streams with intensity functions (Eq. \ref{eq: nest_reply} and \ref{eq: nest_main}), and the parameters can be estimated through maximizing the joint likelihood function since two intensity functions are correlated.

	\subsection{Parameters Estimation and Training Complexity}
	
	Given the complete event sequence history $\mathcal{H}_t$ with associated marks in a point process, the conditional density function can be computed as $f(t) = f(t | \mathcal{H}_t)$:
	\begin{equation}
	\label{eq: eq1}
	\begin{aligned}
	L &= f(t) = f(t_1\cdot t_2 \cdot ... \cdot t_N) = \prod_{n=1}^{N} f\Big[t\rvert t_1, t_2, ..., t_N\Big] \\
	&= \Big[ \prod^{N}_{n=1}\lambda(t_n)\Big] \cdot \exp(-\int^{t_n}_{0}\lambda(u) \diff u)\\
	\end{aligned}
	\end{equation}
	
	Directly computing the likelihood function has the potential risks of underflow, thus it is customary to maximize the log-likelihood function. According to the likelihood function described in Eq. \ref{eq: eq1}, we obtain the log-likelihood function by introducing both main thread stream and reply stream intensity function into Eq. \ref{eq: eq1}. Then, the log-likelihood function is:

	\begin{equation*}
	\begin{aligned}
	\log(L) =& \sum^{m}_{i=1} \sum^{n_i}_{j=1} \Big[\log\big(\lambda_{\main}(t)\big) + \log\big(\lambda_{\reply}^{i}(t)\big)\Big] \\ &- \Lambda_{\main}(t_m) - \sum^{m}_{i=1}\Lambda_{\reply}^{i}(t_{ n_i}),
	\end{aligned}
	\end{equation*}
	where $\Lambda_{\main}(\cdot)$ and $\Lambda_{\reply}^{i}(\cdot)$ is the integrated conditional intensity function, namely the compensator for both main thread stream and associated reply streams \cite{laub2015hawkes}. We present the complete derivation of the log-likelihood function of NesTPP as well as compensators $\Lambda_{\main}(\cdot)$ and $\Lambda_{\reply}^{i}(\cdot)$ in Appendix \ref{appendix:Derivation}.
	
	\begin{algorithm}[t]
		\caption{Adaptive Sampling for Main Threads}\label{algo:nest_main}
		\KwInput{$N$, $M$, $R$, Eq. \ref{eq: nest_reply} and \ref{eq: nest_main}.}
		\KwOutput{The list of sampled main threads time $M$.}
		Set counter $c \gets 0$, $t \gets M[-1]$ (Last item in $M$)\;
		\While{$c$ $< N$}{
			Update reply time series: $R \gets$ Alg. \ref{algo:nest_reply}\;
			Get events intensity upper bound $\lambda_{\main} \gets \lambda_{\main}(t)$ (Eq. \ref{eq: nest_main})\;
			Draw $s \sim U(0, 1)$, and $\tau \gets -\frac{\ln(s)}{\lambda_{\main}}$ \;
			Update the current time: $t \gets t + \tau$ \;
			Draw $s' \sim U(0, \lambda_{\main})$\;
			\eIf{$s' < \lambda_{\main}(t)$}{
				Update the main thread series: $M.$append$(t)$\;
				Update the reply series: $R.$append([ ])\;
				Update the counter: $c \gets c + 1$\;
			}{
				go back to the beginning of the \textbf{while} section\;
			}
		}
	\end{algorithm}

	Due to the non-convexity of this likelihood function, we utilize the limited-memory Broyden-Fletcher-Goldfarb-Shanno algorithm (L-BFGS)\cite{avriel2003nonlinear}, a quasi-Newton gradient-based optimization technique, to solve the nonlinear and unconstrained likelihood function and update parameters. The L-BFGS optimization method has the time complexity $O(k\cdot s)$, where $k$ is the number of parameters that we aim to estimate, and $s$ is the number of steps stored in memory by parameter declaration. The complexity of the log-likelihood function is taken to be $O(\Gamma)$, where the $\Gamma$ is the total number of events in the training set. Taken together, the upper bound of the training complexity of NesTPP is $O(\Gamma \cdot k\cdot s)$.	To avoid the local optimum caused by non-convexity of the likelihood function, we sample several initializations of the parameter set at the start point and choose the parameter set with the best performance. Moreover, we regulate the upper and lower bounds for all the parameters in order to avoid potential calculation overflow and stabilize the training process.  
	
		\begin{algorithm}[t]
		\caption{Adaptive Sampling for Reply Events}\label{algo:nest_reply}
		\KwInput{$T$, $M[i]$, $R$, $C$, Eq. \ref{eq: nest_reply}.}
		\KwOutput{The list of sampled events time $R$.}
		Set counter $c \gets 0$\;
		\For{($i=0; i < len(R); i++$)}{
		    \eIf{$R[i]$ is not empty}{
					$t \gets R[i][-1]$ (Last item in $R[i]$)\;
					}{
					$t \gets M[i]$\;
				}
			\While{$t \le T$ and $c \le C$}{
				Get reply event intensity upper bound $\lambda_{\reply}^{i} \gets \lambda_{\reply}^{i}(t)$ (Eq. \ref{eq: nest_reply})\;
				Draw $s \sim U(0, 1)$, and $\tau \gets -\frac{\ln(s)}{\lambda_{\reply}^{i}}$ \;
				Update the current time: $t \gets t + \tau$ \;
				Draw $s' \sim U(0, \lambda_{\reply}^{i})$\;
				\eIf{$s' < \lambda_{\reply}^{i}(t)$}{
					Update the reply time series: $R[i].$append($t$)\;
					Update the counter: $c \gets c + 1$
				}{
					go back to the head of the \textbf{while} section\;
				}
			}
		}
		
	\end{algorithm}
	\subsection{Properties of NesTPP} \label{subsec: propoerties}
	\textbf{Prediction of Future Dynamics.} Having observed the evolution of an information cascade until the current time under a given TPP model, one can simulate a possible continuation of the cascade using the standard thinning technique \cite{ogata1981lewis}. Through simulating the expected number of all descendent main threads and associated replies, the temporal size of the information cascade can thus be quantified. Considering the natural nested structure of the ODF, we modify the standard thinning technique to adaptively and simultaneously generate main threads and the linked replies. The complete process has been concluded in Alg. \ref{algo:nest_main} and \ref{algo:nest_reply}. 
	
	Specifically, the adaptive sampling algorithm is composed of two parts: simulating main threads and simulating the associated replies of each simulated main thread. The simulation of the main threads takes the expected number of future main threads $N$, the observed main thread stream $M$, the associated reply streams $R$, and the fitted parameters as inputs. At each iteration, the arrival time $t_i$ of a new main thread is estimated through our NesTPP model (Eq. \ref{eq: nest_reply} and \ref{eq: nest_main}), as shown in Alg. \ref{algo:nest_main}. Meanwhile, the arrival time of replies $t_{i, j}$ under each main thread can also be sampled. As stated in Alg. \ref{algo:nest_reply}, the simulation of the reply stream of each main thread takes some extra inputs: $T$, $M[i]$, and $C$. $T$ is the size of the time window in which allows the $i$-th main thread to simulated the associated replies, and $M[i]$ is the arrival time of the linked $i$-th main thread. After the simulated arrival time of a reply exceeds the time window $T$ or the number of the simulated replies exceeds the given threshold $C$, the model would begin to simulate the next $(i+1)$-th main thread. Therefore, the arrival time $t_i$ as well as its linked $t_{i, j}$ of a new main thread can be sampled through our NesTPP model (Eq. \ref{eq: nest_reply} and \ref{eq: nest_main}) simultaneously. 


	\textbf{Final Popularity.} As aforementioned, one can estimate the final cascade size by the standard simulation techniques. However, there also exists a closed-form solution to calculate the expected number of future events in the cascade. Explicitly, as introduced in \cite{laub2015hawkes}, the branching factor $n^*$ of the TPP can be employed to estimate the final cascade size. Through the branching factor, it is also possible to estimate the potential breakout of an information cascade \cite{zhao2015seismic}. We obtain the closed-form expression of the branching factor of the main thread stream by integrating the memory kernel $\psi(t)$ of our main thread stream intensity function (Eq. \ref{eq: nest_main}). 
	
	\newtheorem{corollary}{Corollary}
	
	\begin{corollary}
		The infectious rate $n^*_{\main}$ of the main thread in proposed NesTPP model is the integral of the memory kernel $\psi(\cdot)$ in Eq. \ref{eq: nest_main}.
		
		\begin{equation*}
		\label{eq: branching}
		n^*_{\main} = \int^{\infty}_{0} \psi(t|\mathcal{H}_t) = \sum^{m}_{i=1}\Big[ \frac{\lambda_{\reply}^{i}(t)\cdot p_i^{\gamma}\cdot(c-t_i)^{\eta}}{\eta}\Big]
		\end{equation*}
	\end{corollary}
	
	The branching factor $n^*_{\main}$ describes the expected number of main threads in a complete process, or informally, the virality of the cascade. $n^*_{\main}$ can also indicate whether the information cascade will become an infinite set, i.e., the outbreak of information cascade. For the infectious rate in the subcritical state ($n_{\main}^* < 1$), the size of the information cascade is bounded, which makes it possible to predict the ultimate popularity $R_{\main}^{\infty}$ of the cascade. Conversely, the main thread stream would be viral if the infectious rate is in the supercritical state ($n_{\main}^* > 1$). When $n^*$ is in the subcritical state, the closed-form of the expected number of future thread events $R_{\main}^{\infty}$ is given by:
	\begin{equation*}
	\label{eq: final}
	R_{\main}^{\infty} = \sum^{\infty}_{0} (n_{\main}^*)^i = \frac{1}{1-n_{\main}^*}.
	\end{equation*}
	
	Similarly, one can also identify the infectious rate of reply streams of each main thread by the integral of the reply stream memory kernel function (Eq. \ref{eq: nest_reply}). 
	\begin{figure*}[!t]
		\subfloat[Trend of Main Thread Arrival Time (Sports)]{\label{fig:simu_main}
			\includegraphics[width=0.34\textwidth]{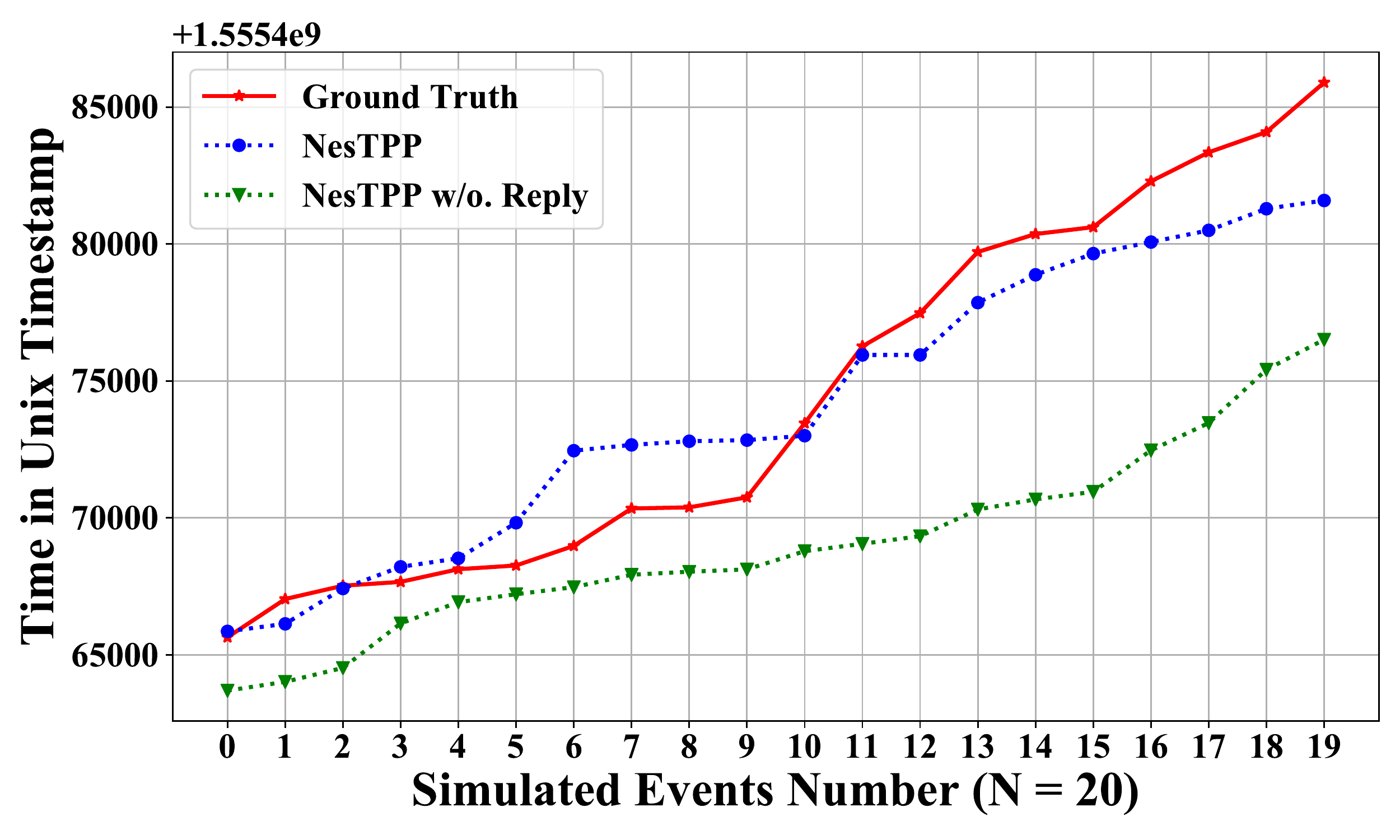}}
		\subfloat[Comparison of Sampled Reply Number (Sports)]{\label{fig:simu_reply_number}\includegraphics[width=0.35\textwidth, height=4cm]{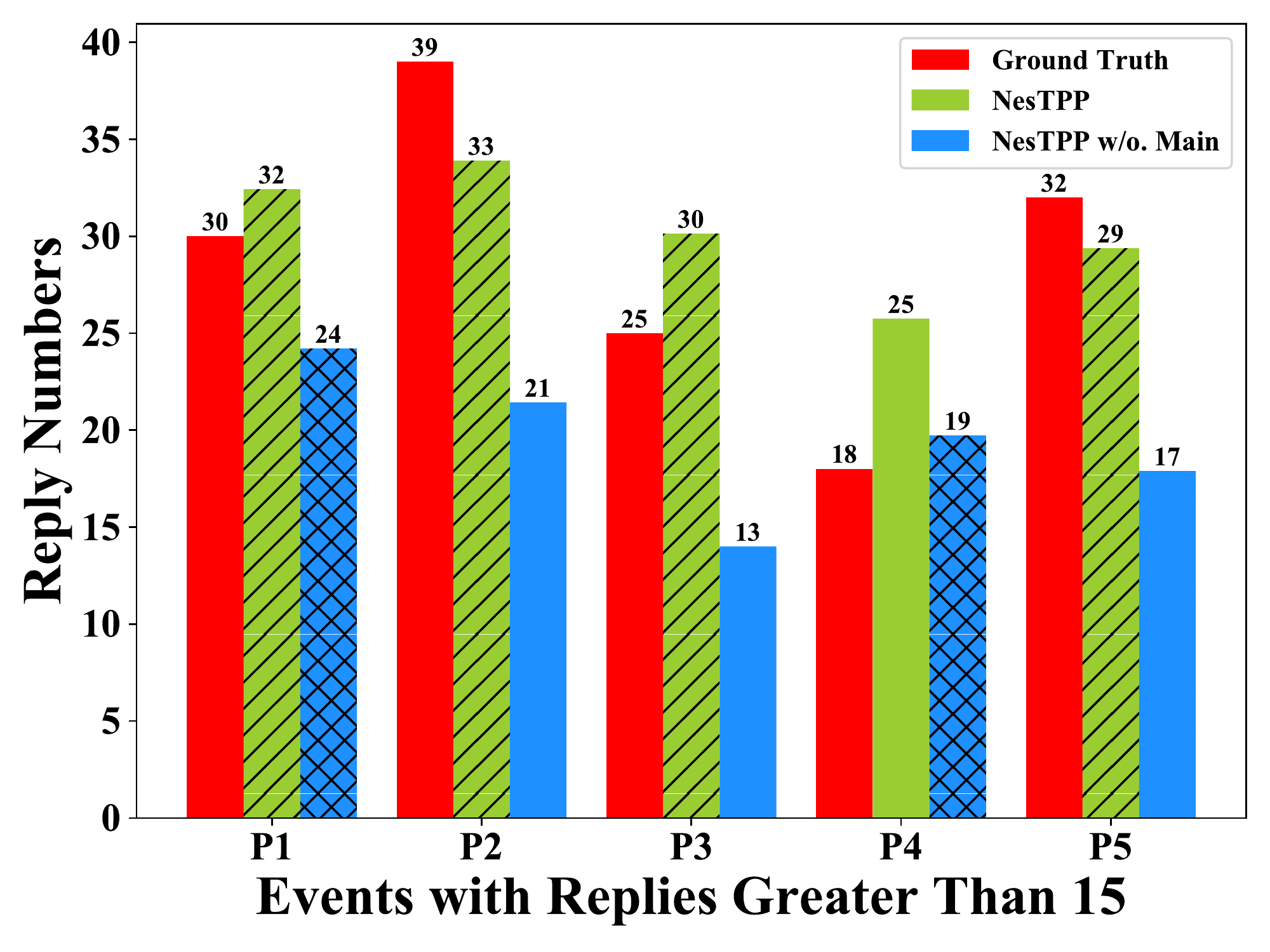}} 
		\subfloat[MAE in Sampled Reply Time (Sports)]{\label{fig:simy_reply_time}\includegraphics[width=0.31\textwidth]{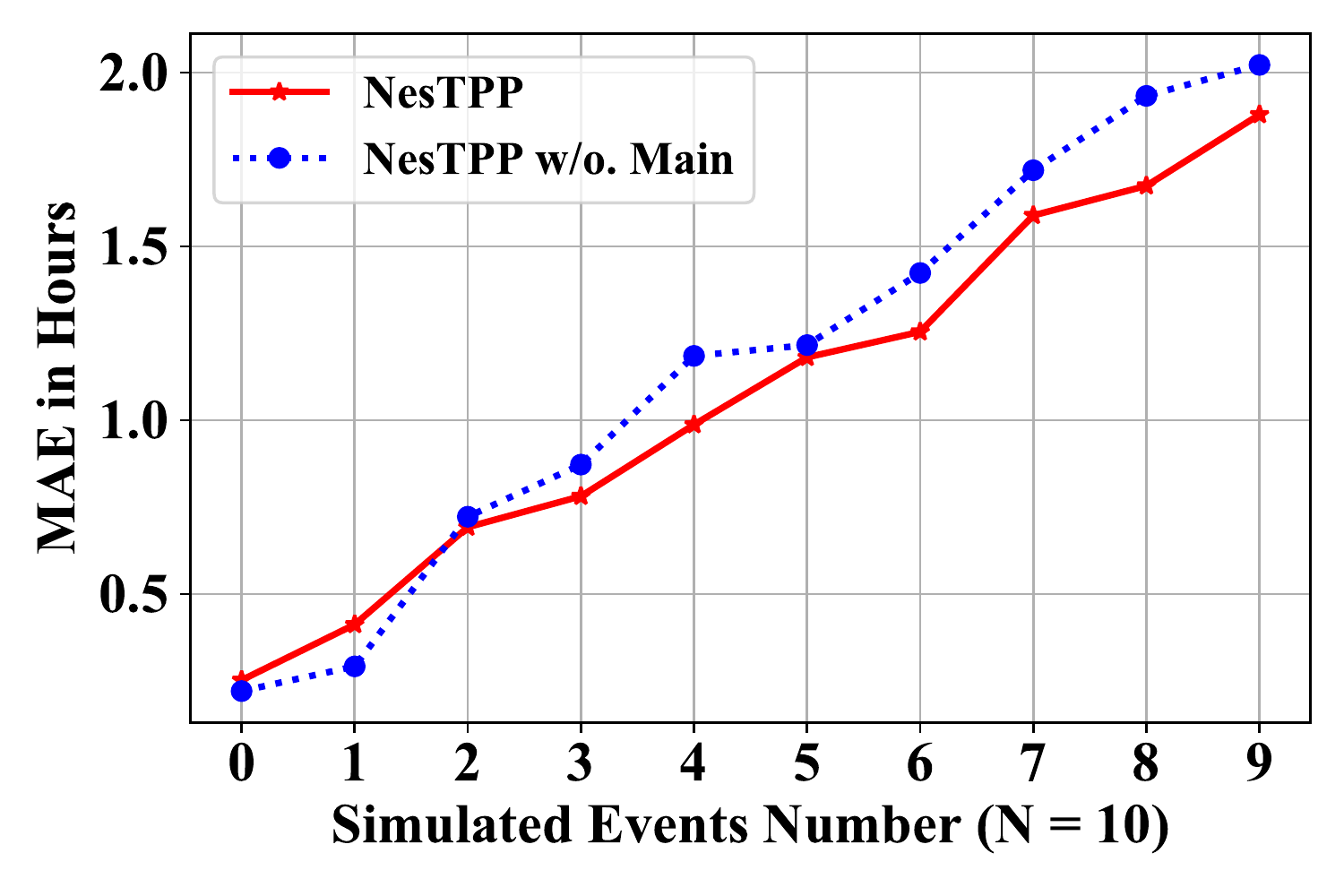}}
		\caption{After training with $1,000$ consecutive main threads and $47,876$ associated replies, we show the performance in different evaluation metrics between NesTPP and NesTPP without the influence of the main thread or reply event stream. In general, the NesTPP model exhibits a better performance in each experiment, which indicates the effectiveness of the correlation between the main thread stream and associated reply stream in modeling information dynamics.}
		\label{fig:distribution}
	\end{figure*}
	
	\section{Experiment}\label{sec: experiment}
	In this section, we present the experiments done for a) examining the effect of the correlation between main threads and associated replies in modeling the cascade dynamics and b) evaluating NesTPP by comparing it with other approaches. In addition, we show that NesTPP, as a visualization tool, can explicitly characterize the correlation between event streams of different types. For reproducibility, we maintain this project on our Github repository\footnote{https://github.com/lingchen0331/NesTPP}.
	
	\subsection{Experiment Setup}
	\begin{table}[t]
		\centering
		\begin{tabular}{@{}c|cc@{}}
			\toprule
			& Sports Dataset & Movie Dataset \\ \midrule 
			Main Threads & 1,610 & 2,261 \\
			Replies & 71,638 & 8,239 \\
			Involved Users & 17,982 & 4,152 \\
			Replies per Thread & 44.496 & 1.98 \\ \bottomrule
		\end{tabular}
		\caption{Dataset Overview}
		\label{tab:dataset}
		\vspace{-2mm}
	\end{table}
	\textbf{Data.} We adopted the dataset of Reddit and selected two popular subforums: Sports and Movie. For the Sports dataset, we collected all the main threads and their replies related to the famous NBA player \textit{LeBron James} during the NBA Playoffs (April 2019), which attracts the most attention in a season. For the movie dataset, we retrieved the discussions of the popular movie \textit{Avengers: Endgame} during the first week since the movie was released. Note that the selection of thread-reply cascades has to be topic-related, since the whole community of an ODF is topic-driven. Table \ref{tab:dataset} gives an overview of two datasets. For both threads and replies, we use UNIX Epoch time to denote the arrival time of events, which is a 10-digit sequence of seconds. Additionally, it can be seen from the table that the distinction in reply density (replies per thread) is obvious, which indicates the main threads in Sports subforum tend to draw more attention than threads in Movie subforum. Moreover, considering the time granularity of both datasets, the amount of the threads in the movie subforum is much more than those in the Sports subforum indicating a different event diffusion pattern. We aim at examining the capability of NesTPP in mining different event patterns with different reply densities.
	
	\textbf{Evaluation Metrics.} We employ the following evaluation metrics to validate the prediction accuracy in our experiment:
	\begin{itemize}
		\item Mean Absolute Error ($\MAE$) in Event Time:
		\begin{equation*}
		{\MAE}_t = \frac{1}{|T|}\sum_{t_i \in T}|t_i - \bar{t}_i|,
		\end{equation*} where $T = \{t_1, t_2, ...\}$ and $\bar{T} = \{\bar{t}_1, \bar{t}_2, ...\}$ are the sequences of ground-truth and simulated event arrival time, respectively. The arrival times $t_i$ and $\bar{t}_i$ in both $T$ and $\bar{T}$ are ordered chronologically.
		
		\item Mean Absolute Error ($\MAE$) in Total Cascade Size:
		\begin{equation*}
		{\MAE}_n = |n - \bar{n}|,
		\end{equation*} where the $n$ and $\bar{n}$ are the ground truth information cascade size (i.e., the total number of simulated replies) and predicted cascade size, respectively.
	\end{itemize}
	
	Furthermore, the simulations of all experiments were repeated until the results converge. We observed that typically $100$ times of simulations would be sufficient for such purpose. During the experiment, we found that the different settings of reply simulation time window $T$ in Alg. \ref{algo:nest_reply} can affect the prediction accuracy in different experiments. In the following experiments, $T$ is set to be $5$ seconds in the event arrival time prediction and $100$ seconds in the total cascade size prediction.

	\subsection{The Correlation Between Main Threads and Reply Posts}\label{sec: correlation}
	In the first experiment, we aim to examine whether or not modeling the nested structure can help better predict the future cascade evolution by comparing NesTPP with regular Hawkes-based models that view each stream as a separate point process. To this end, we modify the intensity function of the main thread stream and reply stream (Eq. \ref{eq: nest_reply} and \ref{eq: nest_main}) assuming the information dynamics of both event streams do not affect each other. The modified conditional intensity functions $\lambda_{\main}'(\cdot)$ and $\lambda_{\reply}'(\cdot)$ are defined by:
	\begin{equation}
	\label{eq: nest_main_modify}
	\lambda_{\main}'(t) = \mu_{\main} + \sum_{i=1}p \cdot (t-t_i + c)^{-(\eta +1)},
	\end{equation} 
	and
	\begin{equation}
	\label{eq: nest_reply_modify}
	\lambda_{\reply}'(t) = \mu_{\reply} + \sum^{n_i}_{j=1} \alpha e^{-\beta(t - t_j)},
	\end{equation}
	where $p \in \mathbb{R}^+$ in Eq. \ref{eq: nest_main_modify} is a parameter which denotes the infectious rate in the power law decaying function \cite{kobayashi2016tideh}. 
	

	\textbf{Influence of Reply Events to Main Threads.} To explore the influence of reply stream in modeling the dynamics of main thread cascade, we fit the NesTPP model with $1,000$ randomly-picked consecutive main threads with associated replies; and we train the control model only with the main threads by the intensity function (Eq. \ref{eq: nest_main_modify}). Next, the two fitted models are applied to simulate the arrival time of the next $20$ threads and compare the simulation results with the arrival time of ground truth threads. The growth trend of both models is shown in Fig. \ref{fig:simu_main}. As shown in the figure, NesTPP consistently outperforms the regular Hawkes-based model. Moreover, there exists a relatively long delay in the arrival time between $9$-th and $11$-th main threads in the figure. As one can see, the NesTPP can infer and predict such a delay by considering the influence of the reply stream,  while the control model fails to capture such subtle changes as the resulted curve tends to be stable all along. 
	
	\textbf{Influence of Main Threads to Reply Events.} We also aim at examining the influence of main threads in predicting the number and arrival time of linked replies. Similarly, we fit NesTPP with $47,876$ reply events as well as the $1,000$ consecutive main threads and simulate the reply stream of immediate $20$ main threads. In the control model, we assume that the reply stream dynamics is governed by a unified point process, and we only fit $47,876$ replies into the intensity function (Eq. \ref{eq: nest_reply_modify}) and simulate the total number of replies for the next $20$ main threads.
	
	We begin by comparing the accuracy of forecasting the number of replies under each main thread, and we focus only on the main threads that have more than $15$ replies for examining the model capability in predicting the large volume replies. Fig. \ref{fig:simu_reply_number} shows the result of the total number of simulated reply events between both experiment models and ground-truth data. On average, the prediction under NesTPP is closer to the ground truth comparing with the one without considering the influence of main thread stream, which demonstrates the positive impact of main thread stream information in forecasting the reply stream dynamics.
	
	Furthermore, we also validate our model in estimating the arrival time of replies under each parent thread. Similarly, we focus on examining the reply reaction time of those threads by the average MAE of the first $10$ replies reaction time among those threads were recorded in Fig. \ref{fig:simy_reply_time}. As the simulated reply number increases, the MAE in time also increases in both models. However, NesTPP-simulated reply time still outperforms the Hawkes-based control model. Therefore, modeling the influence of main thread stream is also effective in predicting the future reply dynamics. Taken together, the nested structure of the ODF and the correlation between the main thread and associated replies are helpful in estimating the future cascade evolution.
	
	\subsection{The Comparison with Existing Approaches} \label{sec: 4c}
	In the second experiment, we assess the performance of NesTPP with other state-of-the-art approaches in terms of predicting the temporal dynamics of information cascade. Note that the selected methods were originally designed for different prediction tasks, and we slightly modify them to fit our data and scenario.
	
	\textbf{Baselines.} 
	We consider three TPP models for comparison, including the non-homogeneous Poisson Process, classic self-exciting point process (Hawkes Process), and the modified SEISMIC model introduced in \cite{zhao2015seismic}. The details of the selected models are as follows:
	\begin{itemize}
		\item \textit{Poisson Process}: The Poisson process is the basic model of the TPP. We adopt the non-homogeneous Poisson process, in which the rate of event arrivals is a function of time, and it can intuitively be interpreted as events are arriving at an average rate of $\lambda$ per unit time.
		
		\item \textit{Hawkes Process}: Hawkes Process has been extensively applied in modeling the temporal dynamics of information cascades in recent literature, and it is one of the most fundamental frameworks adopted in the existing methods (\hspace{1sp}\cite{rizoiu2017expecting, chen2018marked, kobayashi2016tideh, medvedev2018modelling}). We therefore only choose Hawkes Process as one of our baselines. 
		
		\item \textit{SEISMIC}: As one of the state-of-the-art approaches, SEISMIC\cite{zhao2015seismic} has drawn much attention and followed by other recent approaches (\hspace{1sp}\cite{rizoiu2017expecting, du2016recurrent, kobayashi2016tideh, mishra2016feature}). Leveraging Hawkes process, SEISMIC tries to predict the infectivity as well as the final cascade size of a tweet post. We follow the hyper-parameter setting ($c=6.26\times 10^{-4}$, $\theta=0.242$) used in the original paper. Note that we replace the follower's count of a user in SEISMIC to the number of replies for feature consistency with our model. 
	\end{itemize}

	\begin{table}[t]
		\centering
		\resizebox{\columnwidth}{!}
		{\renewcommand{\arraystretch}{1.25}
			\begin{tabular}{@{} c|c|c|c|c|c @{}}
				\toprule
				\multicolumn{2}{c|}{} & Poisson & Hawkes & SEISMIC & NesTPP \\ \midrule
				\multirow{5}{*}{\begin{tabular}[c]{@{}c@{}}Sports\\  Dataset\end{tabular}} & $1$ & 1.694 $\pm$ 1.71 & 0.62 $\pm$ 0.4 & 0.55 $\pm$ 0.3 & \textbf{0.42 $\pm$ 0.51} \\ 
				& $5$ & 1.8 $\pm$ 1.28 & 1.616 $\pm$ 1 & 1.011 $\pm$ 0.88 & \textbf{0.989 $\pm$ 0.86} \\  
				& $10$ & 1.91 $\pm$ 1.29 & 2.949 $\pm$ 1.63 & 1.62 $\pm$ 1.5 & \textbf{1.579 $\pm$ 1.65} \\ 
				& $15$ & 1.852 $\pm$ 1.14 & 4.01 $\pm$ 1.76 & 1.92 $\pm$ 1.67 & \textbf{1.788 $\pm$ 1.26} \\ 
				& $20$ & 2.012 $\pm$ 1.3 & 4.91 $\pm$ 1.84 & 2.08 $\pm$ 1.75 & \textbf{1.838 $\pm$ 1.28} \\ \midrule
				\multirow{5}{*}{\begin{tabular}[c]{@{}c@{}}Movie\\  Dataset\end{tabular}} & $1$ & 1.78 $\pm$ 1.26 & 0.94 $\pm$ 0.89 & 0.78 $\pm$ 0.67 & \textbf{0.22 $\pm$ 0.06} \\  
				& $5$ & 1.84 $\pm$ 1.28 & 1.19 $\pm$ 0.91 & 0.955 $\pm$ 0.47 & \textbf{0.65 $\pm$ 0.07} \\  
				& $10$ & 1.85 $\pm$ 1.22 & 1.146 $\pm$ 0.73 & 1.386 $\pm$ 0.95 & \textbf{1.181 $\pm$ 0.13} \\  
				& $15$ & 1.929 $\pm$ 1.31 & 1.571 $\pm$ 1.12 & 1.552 $\pm$ 1.04 & \textbf{1.536 $\pm$ 0.19} \\  
				& $20$ & 2.089 $\pm$ 1.74 & 2.316 $\pm$ 1.74 & 2.137 $\pm$ 1.45 & \textbf{1.968 $\pm$ 0.27} \\ \bottomrule
		\end{tabular}}
		\caption{Adaptive Prediction of Main Thread Arrival Time}
		\label{tab:main}
	\end{table}
	
	We do not consider other multivariate TPP models since they can only fit and simulate fixed-dimension event sequences. In addition, other non-TPP models either utilized additional mark information or originally applied in other forms of online social networks,  which also cannot be directly employed in our proposed scenario.
	
	\textbf{Prediction of Future Main Thread Stream.} 
	Let us first consider the problem of predicting the arrival time of the future main threads. As described in Sec. \ref{subsec: propoerties}, NesTPP can simulate the arrival time of future main threads using the adaptive thinning technique. Alternatively, since other approaches do not consider the correlation between two event streams, we use standard thinning techniques \cite{ogata1981lewis} for other models to sample the arrival time of future threads. For both datasets, we apply cross-validation to measure the effectiveness of our model and select a total of $10$ groups of $100$ and $150$ randomly-picked consecutive main threads with corresponding replies to simulate the arrival time of the next $20$ main threads in different datasets. We record the average $\MAE_t$ for the first main thread, the first $5$ main threads, the first $10$ main threads, the first $15$ main threads, and the total simulated threads. The result of $\MAE_t$ are summarized in Table \ref{tab:main}. As shown in the table, the average $\MAE_t$ of all models in the sampled threads arrival time increase with time, and NesTPP outperforms other approaches by a lower average error. For other baselines, the simulation of the Poisson process does not incorporate any provided previous main thread history and reply stream information. Therefore, the deviation of the first simulated event arrival time is less accurate than other approaches. However, for this reason, the rate of error cumulation in the adaptive prediction of the Poisson process is slower than other baselines. For other baselines that consider the historical main thread information, Hawkes process and SEISMIC have the close $\MAE_t$ at the first few simulated main threads. The gaps in prediction performance with NesTPP gradually show as the number of simulated main threads grows. With the consideration of the reply stream, the NesTPP model can more accurately predict the temporal evolution of the main threads in terms of arrival time.

	\begin{table}[t]
		\centering
		\resizebox{\columnwidth}{!}
		{\renewcommand{\arraystretch}{1.31}
			\begin{tabular}{@{} c|c|c|c|c|c @{}}
				\toprule
				\multicolumn{2}{c|}{} & Poisson & Hawkes & SEISMIC & NesTPP \\ \midrule
				\multirow{5}{*}{\begin{tabular}[c]{@{}c@{}}Sports\\ Dataset\end{tabular}} & $1$ & 0.059 $\pm$ 0.017 & 0.037 $\pm$ 0.05 & \textbf{0.016 $\pm$ 0.04} & 0.026 $\pm$ 0.015 \\ 
				& $5$ & 0.174 $\pm$ 0.039 & 0.093 $\pm$ 0.07 & 0.065 $\pm$ 0.027 & \textbf{0.049 $\pm$ 0.08} \\  
				& $10$ & 0.313 $\pm$ 0.07 & 0.156 $\pm$ 0.12 & 0.138 $\pm$ 0.11 & \textbf{0.112 $\pm$ 0.073} \\  
				& $20$ & 0.569 $\pm$ 0.17 & 0.441 $\pm$ 0.12 & 0.258 $\pm$ 0.39 & \textbf{0.221 $\pm$ 0.174} \\ 
				& $30$ & 0.843 $\pm$ 0.34 & 0.612 $\pm$ 0.19 & 0.479 $\pm$ 0.7 & \textbf{0.387 $\pm$ 0.456} \\ \midrule
				\multirow{5}{*}{\begin{tabular}[c]{@{}c@{}}Movie\\ Dataset\end{tabular}} & $1$ & 3.755 $\pm$ 1.72 & 1.084 $\pm$ 0.83 & 1.215 $\pm$ 0.46 & \textbf{0.546 $\pm$ 0.84} \\ 
				& $3$ & 5.371 $\pm$ 3.42 & 1.834 $\pm$ 1.24 & 1.923 $\pm$ 1.39 & \textbf{1.057 $\pm$ 0.03} \\ 
				& $6$ & 8.19 $\pm$ 6.28 & 3.49 $\pm$ 1.13 & 2.784 $\pm$ 1.81 & \textbf{1.66 $\pm$ 1.14} \\ 
				& $9$ & 9.75 $\pm$ 6.94 & 4.807 $\pm$ 1.35 & 3.742 $\pm$ 1.94 & \textbf{2.216 $\pm$ 1.35} \\ 
				& $15$ & 12.04 $\pm$ 4.4 & 7.46 $\pm$ 1.8 & 6.32 $\pm$ 2.73 & \textbf{3.34 $\pm$ 1.77} \\ \bottomrule
			\end{tabular}%
		}
		\caption{Adaptive Prediction of Reply Arrival Time}
		\label{tab:reply}
	\end{table}
	
	\textbf{Prediction of Future Reply Stream.} Similarly, following the experiment procedures in sampling the arrival time of the main threads, we now examine the performance of NesTPP in forecasting the reply stream dynamics. In specific, we aim to show the $\MAE_t$ of the simulated reply arrival time of each main thread by given a limited history of each reply stream. Considering the difference between the reply density in different datasets, we choose to test the $\MAE_t$ of different reply numbers in different datasets. We randomly select $100$ and $50$ reply streams and provide all models the beginning $10$ reply arrival time of each reply stream in the Sports dataset and Movie dataset, respectively. We record the relative error of sampling a total of $30$ and $15$ subsequent replies in both datasets, and the results are shown in Table \ref{tab:reply}. Overall, by modeling the reply event infectivity in Eq. \ref{eq: nest_reply}, NesTPP provides a consistent improvement in predicting the long-term reply stream diffusion dynamics comparing with other methods by an evident margin in both datasets. Although the $\MAE_t$ of the first predicted replies in the Sports dataset are close among all baselines and the performance of NesTPP is not the best, it is needed to consider the high randomness of predicting a single event in the reply stream.

	\textbf{Total Cascade Popularity.} In this experiment, we aim to test the model's ability to predict the final cascade size (i.e., the total number of replies for all simulated main threads). We apply the adaptive sampling method to simulate the main threads and associated replies together. For other baselines, we firstly simulate the total number of main threads up to the pre-defined time limit, we then sample reply streams for each simulated main thread with the time limit up to the arrival time of the last simulated main thread. On both datasets, we apply $10$-fold cross-validation to test the relative error $\MAE_n$ between the prediction result and ground truth on the randomly-picked time point's following time durations. Based on the different time granularity in both datasets, we choose to test the prediction performance with different window sizes. The results are summarized in Table \ref{tab:popularity_sport} and \ref{tab:popularity_movie}.
	\begin{table}[t]
		\centering
			{\renewcommand{\arraystretch}{1.24}
		\resizebox{0.9\columnwidth}{!}{%
		\begin{tabular}{@{}c|ccccc@{}}
\toprule
\multirow{2}{*}{Model} & \multicolumn{5}{c}{Sports Dataset} \\ \cmidrule(l){2-6} 
 & $0.5$ Hrs & $1$ Hrs & $3$ Hrs & $9$ Hrs & $12$ Hrs \\ \midrule
Poisson & 64.09 & 108.4 & 98.52 & 364.17 & 468.5 \\
Hawkes & 63.89 & 109.36 & 114.63 & 380.1 & 459.26 \\
SEISMIC & 60.91 & 98.64 & 101.12 & 316.03 & 425.75 \\
NesTPP & \textbf{56.61} & \textbf{87.85} & \textbf{95.25} & \textbf{239.81} & \textbf{400.75} \\ \bottomrule
\end{tabular}%
		}}
		\caption{Total Cascade Size Prediction of Sports Dataset}
		\label{tab:popularity_sport}
	\end{table}
	
     We can see from both tables that the average prediction errors for all models gradually increases with the enlarging prediction window size because of the accumulated bias exposure when conducting long-range prediction. In the Sports dataset, comparing with other non-homogeneous TPP, NesTPP can consistently provide a more reliable prediction result in different window sizes, and the margin becomes more evident as the time window size increases. Similarly, the relative prediction error of NesTPP in the Movie dataset also exhibits a clear prediction gap with other approaches. Considering the reply density varies in both datasets, NesTPP shows excellent robustness when forecasting the ODF temporal cascade size in different reply intensities. By modeling the nested structure and mutual influence between two event streams, NesTPP can capture the arrival frequency of replies and provide a more reliable prediction result.
    
	
	In summary, NesTPP shows a competitive results in different prediction tasks comparing with other state-of-the-art approaches in most cases. By modeling the natural nested structure with explicit intensity functions, NesTPP can be better served at characterizing the cascade dynamics in ODFs operated in the thread-reply mode. 
	
	\subsection{Visualization of Correlation Between Cascades}
\begin{figure}[tbp]
		\centerline{\includegraphics[width=0.49\textwidth]{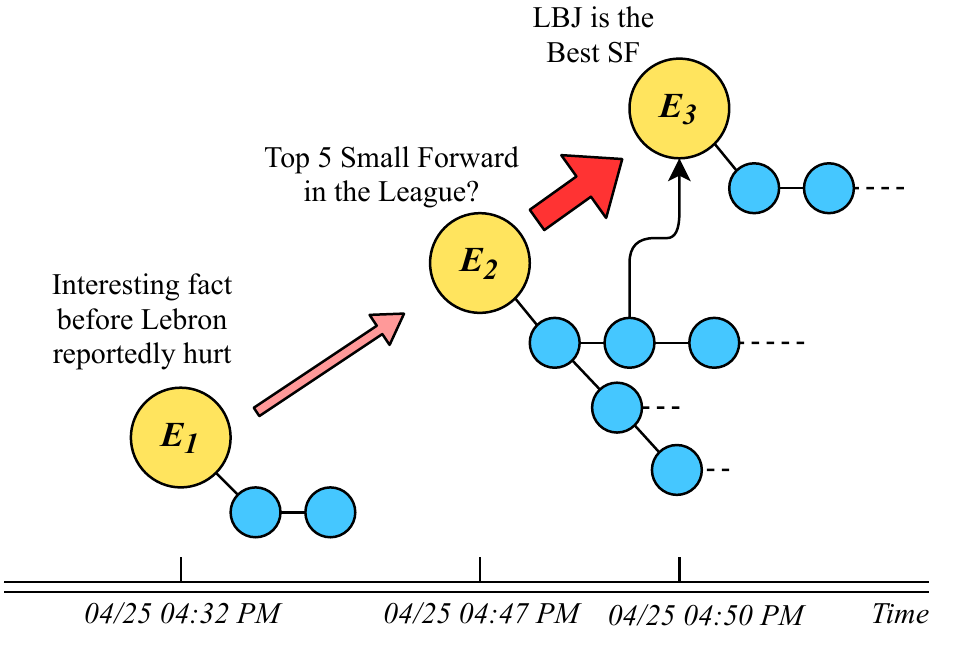}}
		\caption{Correlation visualization between cascades: The yellow node $E_i$ represents the main threads occurs in a non-negative timeline, and the associated blue nodes linked to a main thread denote the replies to the main thread.}
		\label{fig:case_study}
	\end{figure}

	In addition to the previous applications, NesTPP can be utilized as a tool to visualize the correlation between cascades. Through analyzing the value of the reply stream intensity in main thread intensity function (Eq. \ref{eq: nest_main}), one can determine the influence of the linked reply stream in modeling the main thread dynamics, so that the correlation between thread-reply cascade can thus be characterized. For a case study, we select three consecutive main threads $E_1$, $E_2$, and $E_3$ discussing the famous basketball player - LeBron James during the NBA Playoffs in the Sports dataset, where each main thread is connected with a reply stream.  In Fig. \ref{fig:case_study}, we visually present such correlations according to the intensity values. Thread $E_1$ has relatively fewer replies than others since users show less interest in the discussion topic, which illustrates that the non-topical thread fails to stimulate the arrival of a new thread with the similar content. After $15$ minutes, the next thread debating a controversial topic of the player draws tremendous attention and rapidly generates a conversation wave. While the thread $E_2$ is still under discussion, a new main thread $E_3$ is posted after $3$ minutes, which can be viewed as the continuity of the conversation in thread $E_2$. More interestingly, the person who posts $E_3$ is observed to be one of the repliers under the thread $E_2$, which can be interpreted as the discussion rush of a controversial topic can rapidly spur the occurrence of a related thread in an ODF; and followers in the related threads may also turn to start a new thread discussing the similar content.
	\begin{table}[t]
		\centering
			{\renewcommand{\arraystretch}{1.2}
		\resizebox{0.9\columnwidth}{!}{%
		\begin{tabular}{@{}c|ccccc@{}}
        \toprule
        \multirow{2}{*}{Model} & \multicolumn{5}{c}{Movie Dataset} \\ \cmidrule(l){2-6} 
        & $0.5$ Hrs & $1$ Hrs & $3$ Hrs & $4.5$ Hrs & $6$ Hrs \\ \midrule
        Poisson & 11.99 & 26.45 & 40.29 & 102.9 & 223.21 \\
        Hawkes & 13.02 & 19.01 & 68.11 & 78.69 & 192.71 \\
        SEISMIC & 16.78 & 21.43 & 45.51 & 68.54 & 132.94 \\
        NesTPP & \textbf{6.99} & \textbf{10.13} & \textbf{33.61} & \textbf{56.75} & \textbf{96.35} \\ \bottomrule
\end{tabular}%
		}}
		\caption{Total Cascade Size Prediction of Movie Dataset}
		\label{tab:popularity_movie}
	\end{table} 
	
	\section{Conclusion and Future Work} \label{sec: conclusion}
	\textbf{Conclusion.} In this paper, by utilizing the self-exciting point process, we have introduced a novel framework - NesTPP that models the correlation between different types of event streams in the ODF using the arrival time of each event and self-contained mark information. Further, we have derived a maximum likelihood approach to estimate parameters in the proposed model. Two individual but correlated sampling methods are proposed to simulate the arrival of the future main threads and associated replies simultaneously. Through the competitive experiment results and additional applications, we verify the NesTPP is not only theoretically grounded but also effective in most cases.
	
	\textbf{Future Work.} 
	Firstly, NesTPP can easily be transferred to other social network scenarios with the thread-reply communication structure. In addition, online users may reply to one of the replies in a cascade instead of directly replying to the main thread, which forms the so-called \textit{nest-of-nest} structure. Modeling the nest-of-nest structure may capture more subtle changes in the temporal dynamics of the reply stream. Last but not least, recurrent neural networks (RNN) are prominent in modeling the complex sequential dependencies, and a few successful examples (\hspace{1sp}\cite{mei2017neural, du2016recurrent, cao2017deephawkes}) have applied the RNN architectures to replace the intensity function in the point process. One concrete future work is to employ multi-dimensional RNN to model the intricate nested communication structure in ODFs.
	
	\appendix
	\section{Proofs}
	\label{appendix: a}

	\subsection{Likelihood Function Derivation}
	\label{appendix:Derivation}

	
	 Given the finite set of the main thread stream arrival time up to but not including time $t$:  $\mathcal{H}_t=\{t_{1},...,t_{m}\}$, and the set of the arrival time of replies of $i$-th main thread $\mathcal{H}(t_{i, j})=\{t_{i, 1}, t_{i, 2}, ..., t_{i, n_i}\}$ on the same timeline. Let $f(t)$ be the joint conditional density function of the time of the next event given the history of all previous events, 
	\begin{equation}
	\label{eq: pdf}
	\begin{aligned}
	f(t | \mathcal{H}(t_i), \mathcal{H}(t_{i, j})) &= 
	f(t_{1}, t_{1, 1}, t_2, t_{2, 1}..., t_{m, 1}, ...) \\
	&= \prod^{m}_{i=1}f(t_{i} | \mathcal{H}_{t_{i}})\prod^{n_i}_{j=1}f(t_{i, j} | \mathcal{H}_{t_{i, j}}), 
	\end{aligned}
	\end{equation} which considers the historical occurrence of all previous events from both main and reply stream. $m$ denotes the total number of posts from main stream, and $n_i$ denotes the number of the reply stream events of the $i$-th main thread. 
	
	According to the past literature \cite{rizoiu2017hawkes}, the event intensity $\lambda(t)$ for both event streams can be expressed in terms of the conditional density $f(t)$ and its corresponding cumulative distribution function $F(t)$:
	\begin{equation}
	\label{eq: intensity}
	\lambda(t) = \frac{f(t)}{1 - F(t)} = \frac{\frac{\partial}{ \partial t}F(t)}{1 - F(t)} = - \frac{\partial}{\partial t} \log{(1 - F(t))}, 
	\end{equation} 
	where the $\lambda(t)$ represents the conditional intensity function of main stream events as we stated in Eq. \ref{eq: nest_main}. For main stream posts, the conditional intensity function can be derived as integrating both sides of Eq. \ref{eq: intensity} over the interval $(t_m, t)$:
	\begin{equation}
	\label{eq: integrate}
	\begin{aligned}
	\int^{t}_{t_m}\lambda_{\main}(u) \diff u &= \log(1 - F_{\main}(t_m))} - \log{(1 - F_{\main}(t))\\
	&= -\log(1 - F_{\main}(t)).
	\end{aligned}
	\end{equation}	
	
	Rearranging the Eq. \ref{eq: integrate}, we can get the expression of cumulative density function $F_{\main}(t)$, and joint probability density function $f_{\main}(t)$ for main stream posts.
	\begin{equation*}
	\label{eq: cdf_main}
	\begin{aligned}
	&F_{\main}(t) = 1 - \exp(-\int^{t}_{t_m}\lambda_{\main}(u) \diff u)\\
	&f_{\main}(t) = \lambda_{\main}(t)\cdot \exp(-\int^{t}_{t_m} \lambda_{\main}(u) \diff u)
	\end{aligned}
	\end{equation*}	
	Similarly, we can get the expression of joint likelihood function for the reply stream of the $i$-th main stream: 
	\begin{equation*}
	\label{eq: pdf_reply}
	\begin{aligned}
	f_{\reply}^i(t) = \lambda_{\reply}^{i}(t)\cdot \exp(-\int^{t}_{t_{m, n_i}}\lambda_{\reply}^{i}(u) \diff u)
	\end{aligned}
	\end{equation*}	
	
	By introducing the reply thread stream and main stream intensity function (Eq. \ref{eq: nest_reply}, \ref{eq: nest_main}) into the likelihood function Eq. \ref{eq: pdf}, the complete likelihood function $L$ can be derived as:
	\begin{equation}
	\label{eq: likelihood_function_app}
	\begin{aligned}
	L &= \prod_{i=1}^{m}[f_{\main}(t_i|\mathcal{H}_{t_i})]\cdot \prod_{j=1}^{n_i}[f^{i}_{\reply}(t_{i, j}|\mathcal{H}_{t_{i, j}})]\\
	&= \prod_{i=1}^{m}\Big[\lambda_{\main}(t_i)\cdot \exp(-\int_{t_{i-1}}^{t_{i}}\lambda_{\main}(u)\diff u)\\
	&\hspace{1em} \cdot \prod_{j=1}^{n_i}\lambda_{\reply}^i(t_i)\cdot \exp(\int_{t_{i, j-1}}^{t_{i, j}}\lambda_{\reply}^i(u)\diff u)\Big]\\
	&= \Big[\prod_{i=1}^{m}\prod_{j=1}^{n_i}\lambda_{\main}(t_i)\cdot \lambda_{\reply}^i(t_{i, j})\Big]\\
	&\hspace{1em} \cdot \prod_{i=1}^{m}\exp(-\int_{t_{n_1}}^{t_{n_i}}\lambda_{\reply}^i(u)\diff u) \cdot \exp(-\int_{0}^{t_m}\lambda_{\main}(u)\diff u).
	\end{aligned}
	\end{equation}
	
	From the likelihood function Eq. \ref{eq: likelihood_function_app}, we can estimate the set of parameters of our NesTPP model by maximizing the likelihood function. To avoid the potential underflow of calculating the very small likelihood, we instead maximize the log of the likelihood function:
	
	\begin{equation}
	\label{eq: log_likelihood_function_app}
	\begin{aligned}
	l &= \log(L) = \Big[\sum_{i=1}^{m}\sum_{j=1}^{n_i}\log(\lambda_{\main}(t_i))+\log(\lambda_{\reply}^i(t_{i, j}))\Big]\\
	&\hspace{1em} - \Lambda_{\main}(t_m) - \sum_{i=1}^{m}\Lambda_{\reply}^i(t_{n_i}),
	\end{aligned}
	\end{equation}
	where the $\Lambda_{\main}(t_m)$ and $\Lambda_{\reply}^{i}(t_{n_i})$ in Eq. \ref{eq: log_likelihood_function_app} represent the compensator of the main and reply event streams. We also present the derivation of both compensators here. 
	
	\begin{equation}
	\label{eq: main_compensator}
	\begin{aligned}
	\Lambda_{\main}(t_m) &= \int_{0}^{t_m}\lambda_{\main}(u)\diff u \\
	&= \int_{0}^{t_m}\mu_{\main}+\psi_{\main}(u, t_m)\diff u \\
	&= \int_{0}^{t_m}\mu_{\main}+ \sum_{0}^{m-1}\int_{i}^{i+1}\psi_{\main}(u, t_m)\diff u\\
	&= \mu_{\main}t_m + \sum_{i=1}^{m}\int_{t_i}^{t_m}\psi_{\main}(u, t_i)\diff u.
	\end{aligned} 	
	\end{equation}
	By introducing Eq. \ref{eq: nest_main} into Eq. \ref{eq: main_compensator}, we have the complete equation of main stream compensator $\Lambda_{\main}(\cdot)$.
	
	\begin{equation}
	\label{eq: main_compensator_2}
	\begin{aligned}
	\Lambda_{\main}(t_m) &= \sum_{i=1}^{m}\int_{t_i}^{t_m} \lambda_{\reply}^i(t_i)(p_i)^{\gamma}(u-t_i+c)^{-(1+\eta)}\diff u\\
	& \hspace{1em} + \mu_{\main}t_m \\
	&= \sum_{i=1}^{m}\Big[\mu_{\reply}+q(t_m)\cdot \sum_{t_{i, j}<t_i}\alpha e^{-\beta(t_i-t_{i, j})}\Big]\\
	&\hspace{1em} \cdot (p_i)^{\gamma}\Big[\frac{1}{\eta c^{\eta}} - \frac{(t_m-t_i+c)^{-\eta}}{\eta}\Big]+\mu_{\main}t_m.
	\end{aligned} 	
	\end{equation}
	
	Similarly, we can derive the reply stream compensator $\Lambda_{\reply}^i(\cdot)$.
	\begin{equation}
	\label{eq: reply_compensator}
	\begin{aligned}
	\Lambda_{\reply}^i(t_{n_i}) &= \int_{0}^{t_1}\mu_{\reply}\diff u + \sum_{j=1}^{n_i-1}\int_{t_j}^{t_{j+1}}\mu_{\reply}\\
	&\hspace{1em} +q(u)\cdot\sum_{t_{k}<u}\alpha e^{-\beta(u-t_k)}\diff u\\
	&= \mu_{\reply}t_{n_i}\\
	&\hspace{1em} +q(t_{n_m-1})\cdot \alpha \Big[ \sum_{j=1}^{n_m-1}\int_{t_{j+1}}^{t_j}\sum_{k=1}^{j}e^{-\beta(u-t_k)}u\Big]\\
	&= \mu_{\reply}t_{n_i} - \frac{\alpha}{\beta}\cdot q(t_{n_m})\sum_{j=1}^{n_m}\Big[e^{-\beta(t_{n_i}-t_j)}-1\Big].
	\end{aligned} 
	\end{equation}
	
	Finally, we have the reply stream and main thread stream intensity functions (Eq. \ref{eq: nest_reply} and \ref{eq: nest_main}) as well as the main thread stream and reply event stream compensators (Eq. \ref{eq: main_compensator} and \ref{eq: reply_compensator}). The complete log-likelihood function that we aim to maximize can be derived from introducing the above equations into the log-likelihood function Eq. \ref{eq: log_likelihood_function_app}.

	\bibliographystyle{IEEEtran}
	\bibliography{clingbib}

\begin{thebibliography}{10}
\providecommand{\url}[1]{#1}
\csname url@samestyle\endcsname
\providecommand{\newblock}{\relax}
\providecommand{\bibinfo}[2]{#2}
\providecommand{\BIBentrySTDinterwordspacing}{\spaceskip=0pt\relax}
\providecommand{\BIBentryALTinterwordstretchfactor}{4}
\providecommand{\BIBentryALTinterwordspacing}{\spaceskip=\fontdimen2\font plus
\BIBentryALTinterwordstretchfactor\fontdimen3\font minus
  \fontdimen4\font\relax}
\providecommand{\BIBforeignlanguage}[2]{{%
\expandafter\ifx\csname l@#1\endcsname\relax
\typeout{** WARNING: IEEEtran.bst: No hyphenation pattern has been}%
\typeout{** loaded for the language `#1'. Using the pattern for}%
\typeout{** the default language instead.}%
\else
\language=\csname l@#1\endcsname
\fi
#2}}
\providecommand{\BIBdecl}{\relax}
\BIBdecl

\bibitem{smedley2018practical}
R.~M. Smedley and N.~S. Coulson, ``A practical guide to analysing online
  support forums,'' \emph{Qualitative Research in Psychology}, pp. 1--28, 2018.

\bibitem{asur2010predicting}
S.~Asur and B.~A. Huberman, ``Predicting the future with social media,'' in
  \emph{Proc. of WI-IAT}, 2010, pp. 492--499.

\bibitem{lan2018personalized}
A.~S. Lan, J.~C. Spencer, Z.~Chen, C.~G. Brinton, and M.~Chiang, ``Personalized
  thread recommendation for mooc discussion forums,'' in \emph{Proc. of ECML
  PKDD}, 2018, pp. 725--740.

\bibitem{wu2016mining}
L.~Wu, F.~Morstatter, X.~Hu, and H.~Liu, ``Mining misinformation in social
  media,'' in \emph{Big Data in Complex and Social Networks}.\hskip 1em plus
  0.5em minus 0.4em\relax Chapman and Hall/CRC, 2016, pp. 135--162.

\bibitem{rizoiu2017expecting}
M.-A. Rizoiu, L.~Xie, S.~Sanner, M.~Cebrian, H.~Yu, and P.~Van~Hentenryck,
  ``Expecting to be hip: Hawkes intensity processes for social media
  popularity,'' in \emph{Proc. of WWW}, 2017, pp. 735--744.

\bibitem{liu2019latent}
S.~Liu, S.~Yao, D.~Liu, H.~Shao, Y.~Zhao, X.~Fu, and T.~Abdelzaher, ``A latent
  hawkes process model for event clustering and temporal dynamics learning with
  applications in github,'' in \emph{Proc. of ICDCS}.\hskip 1em plus 0.5em
  minus 0.4em\relax IEEE, 2019, pp. 1275--1285.

\bibitem{im2006online}
E.-O. Im and W.~Chee, ``An online forum as a qualitative research method:
  practical issues,'' \emph{Nursing research}, vol.~55, no.~4, p. 267, 2006.

\bibitem{zhao2015seismic}
Q.~Zhao, M.~A. Erdogdu, H.~Y. He, A.~Rajaraman, and J.~Leskovec, ``Seismic: A
  self-exciting point process model for predicting tweet popularity,'' in
  \emph{Proc. of SIGKDD}, 2015.

\bibitem{kobayashi2016tideh}
R.~Kobayashi and R.~Lambiotte, ``Tideh: Time-dependent hawkes process for
  predicting retweet dynamics,'' in \emph{Proc. of ICWSM}, 2016.

\bibitem{mei2017neural}
H.~Mei and J.~M. Eisner, ``The neural hawkes process: A neurally
  self-modulating multivariate point process,'' in \emph{Proc. of NIPS}, 2017,
  pp. 6754--6764.

\bibitem{chen2018marked}
F.~Chen and T.~Hong, ``Marked self-exciting point process modelling of
  information diffusion on twitter,'' \emph{The Annals of Applied Statistics},
  vol.~12, no.~4, pp. 2175--2196, 2018.

\bibitem{cheng2014can}
J.~Cheng, L.~Adamic, P.~A. Dow, J.~M. Kleinberg, and J.~Leskovec, ``Can
  cascades be predicted?'' in \emph{Proc. of WWW}, 2014, pp. 925--936.

\bibitem{thomas2002learning}
M.~J. Thomas, ``Learning within incoherent structures: The space of online
  discussion forums,'' \emph{Journal of Computer Assisted Learning}, vol.~18,
  no.~3, pp. 351--366, 2002.

\bibitem{aumayr2011reconstruction}
E.~Aumayr, J.~Chan, and C.~Hayes, ``Reconstruction of threaded conversations in
  online discussion forums,'' in \emph{Proc. of AAAI ICWSM}, 2011.

\bibitem{medvedev2018modelling}
A.~N. Medvedev, J.-C. Delvenne, and R.~Lambiotte, ``Modelling structure and
  predicting dynamics of discussion threads in online boards,'' \emph{Journal
  of Complex Networks}, vol.~7, no.~1, pp. 67--82, 2018.

\bibitem{hawkes1971point}
A.~G. Hawkes, ``Point spectra of some mutually exciting point processes,''
  \emph{Journal of the Royal Statistical Society: Series B (Methodological)},
  vol.~33, no.~3, pp. 438--443, 1971.

\bibitem{iwata2013discovering}
T.~Iwata, A.~Shah, and Z.~Ghahramani, ``Discovering latent influence in online
  social activities via shared cascade poisson processes,'' in \emph{Proc. of
  SIGKDD}, 2013, pp. 266--274.

\bibitem{linderman2014discovering}
S.~Linderman and R.~Adams, ``Discovering latent network structure in point
  process data,'' in \emph{Proc. ICML}, 2014, pp. 1413--1421.

\bibitem{blundell2012modelling}
C.~Blundell, J.~Beck, and K.~A. Heller, ``Modelling reciprocating relationships
  with hawkes processes,'' in \emph{Advances in Neural Information Processing
  Systems}, 2012, pp. 2600--2608.

\bibitem{cao2017deephawkes}
Q.~Cao, H.~Shen, K.~Cen, W.~Ouyang, and X.~Cheng, ``Deephawkes: Bridging the
  gap between prediction and understanding of information cascades,'' in
  \emph{Proc. of CIKM}, 2017.

\bibitem{yang2003effects}
H.-L. Yang and J.-H. Tang, ``Effects of social network on students'
  performance: a web-based forum study in taiwan,'' \emph{Journal of
  Asynchronous Learning Networks}, vol.~7, no.~3, pp. 93--107, 2003.

\bibitem{dawson2008study}
S.~Dawson, ``A study of the relationship between student social networks and
  sense of community,'' \emph{Journal of educational technology \& society},
  vol.~11, no.~3, pp. 224--238, 2008.

\bibitem{yusof2009students}
N.~Yusof, A.~A. Rahman \emph{et~al.}, ``Students' interactions in online
  asynchronous discussion forum: A social network analysis,'' in \emph{2009
  International Conference on Education Technology and Computer}.\hskip 1em
  plus 0.5em minus 0.4em\relax IEEE, 2009, pp. 25--29.

\bibitem{gomez2011modeling}
V.~G{\'o}mez, H.~J. Kappen, and A.~Kaltenbrunner, ``Modeling the structure and
  evolution of discussion cascades,'' in \emph{Proc. of HT}, 2011, pp.
  181--190.

\bibitem{wang2012user}
C.~Wang, M.~Ye, and B.~A. Huberman, ``From user comments to on-line
  conversations,'' in \emph{Proc. of SIGKDD}, 2012, pp. 244--252.

\bibitem{nishi2016reply}
R.~Nishi, T.~Takaguchi, K.~Oka, T.~Maehara, M.~Toyoda, K.-i. Kawarabayashi, and
  N.~Masuda, ``Reply trees in twitter: data analysis and branching process
  models,'' \emph{Social Network Analysis and Mining}, vol.~6, no.~1, p.~26,
  2016.

\bibitem{bakshy2011everyone}
E.~Bakshy, J.~M. Hofman, W.~A. Mason, and D.~J. Watts, ``Everyone's an
  influencer: quantifying influence on twitter,'' in \emph{Proc. of WSDM},
  2011, pp. 65--74.

\bibitem{naveed2011bad}
N.~Naveed, T.~Gottron, J.~Kunegis, and A.~C. Alhadi, ``Bad news travel fast: A
  content-based analysis of interestingness on twitter,'' in \emph{Proc. of
  WWW}.\hskip 1em plus 0.5em minus 0.4em\relax ACM, 2011, p.~8.

\bibitem{zaman2010predicting}
T.~R. Zaman, R.~Herbrich, J.~Van~Gael, and D.~Stern, ``Predicting information
  spreading in twitter,'' in \emph{Workshop on computational social science and
  the wisdom of crowds, nips}, vol. 104, no.~45, 2010, pp. 17\,599--601.

\bibitem{rizoiu2017hawkes}
M.-A. Rizoiu, Y.~Lee, S.~Mishra, and L.~Xie, ``Hawkes processes for events in
  social media,'' in \emph{Frontiers of Multimedia Research}, 2017, pp.
  191--218.

\bibitem{ozaki1979maximum}
T.~Ozaki, ``Maximum likelihood estimation of hawkes' self-exciting point
  processes,'' \emph{Annals of the Institute of Statistical Mathematics},
  vol.~31, no.~1, pp. 145--155, 1979.

\bibitem{laub2015hawkes}
P.~J. Laub, T.~Taimre, and P.~K. Pollett, ``Hawkes processes,'' \emph{arXiv
  preprint arXiv:1507.02822}, 2015.

\bibitem{avriel2003nonlinear}
M.~Avriel, \emph{Nonlinear programming: analysis and methods}.\hskip 1em plus
  0.5em minus 0.4em\relax Courier Corporation, 2003.

\bibitem{ogata1981lewis}
Y.~Ogata, ``On lewis' simulation method for point processes,'' \emph{ITIT},
  vol.~27, no.~1, pp. 23--31, 1981.

\bibitem{du2016recurrent}
N.~Du, H.~Dai, R.~Trivedi, U.~Upadhyay, M.~Gomez-Rodriguez, and L.~Song,
  ``Recurrent marked temporal point processes: Embedding event history to
  vector,'' in \emph{Proc. of SIGKDD}, 2016, pp. 1555--1564.

\bibitem{mishra2016feature}
S.~Mishra, M.-A. Rizoiu, and L.~Xie, ``Feature driven and point process
  approaches for popularity prediction,'' in \emph{Proc. of CIKM}, 2016.

\end{thebibliography}
	
\end{document}